\documentstyle[preprint,eqsecnum,aps,pra]{revtex}
\tighten
\begin{document}
\draft

\title{Electrostatic Equilibrium of Two Spherical \\
Charged Masses in General Relativity}
\author{G.~P.~Perry and F.~I.~Cooperstock}
\address{Department of Physics and Astronomy, University 
of Victoria \\
P.O. Box 3055, Victoria, B.C. V8W 3P6 (Canada)}
\date{\today}
\maketitle   

\begin{abstract}
Approximate solutions representing the gravitational-electrostatic
balance of two arbitrary point sources in general relativity have led
to contradictory arguments in the literature with respect to the
condition of balance. Up to the present time, the only known exact
solutions which can be interpreted as the non-linear superposition of
two spherically symmetric (Reissner-Nordstr\"{o}m) bodies without an
intervening strut has been for critically charged masses, $M^2_i =
Q^2_i$.  In the present paper, an exact electrostatic solution of the
Einstein-Maxwell equations representing the exterior field of two
arbitrary charged Reissner-Nordstr\"{o}m bodies in equilibrium is
studied. The invariant physical charge for each source is found by
direct integration of Maxwell's equations. The physical mass for each
source is invariantly defined in a manner similar to which the charge
was found. It is shown through numerical methods that balance without
tension or strut can occur for non-critically charged bodies.  It is
demonstrated that other authors have not identified the correct
physical parameters for the mass and charge of the sources. Further
properties of the solution, including the multipole structure and
comparison with other parameterizations, are examined.
\end{abstract}
\pacs{04.20.Jb,~04.40.Nr}

\section{Introduction}
In a recent paper by Bonnor \cite{bonnor1}, the equilibrium conditions
for a charged test particle in the field of a spherically symmetric
charged mass (Reissner-Nordstr\"{o}m solution) were investigated. He
found that the classical condition for equilibrium
\begin{equation} \label{classical} 
M_1 M_2 = Q_1 Q_2 
\end{equation} 
for which the separation between the particles is arbitrary, was
neither necessary nor sufficient for electrostatic balance of two
spherical masses. This is in conflict with the earlier results of
Barker and O'Connell \cite{barker} and Ohta and Kimura \cite{ohta} who
used different approximation methods.  Barker and O'Connell claimed
that in the post-Newtonian approximation, the equation
\begin{equation} 
(M_1 Q_2 - M_2 Q_1)(Q_1-Q_2) = 0  
\end{equation}  
had to be satisfied in addition to (\ref{classical}). Ohta and Kimura
claimed that in the post-post-Newtonian approximation, the necessary
and sufficient condition for balance is that each mass be ``critically''
charged, 
\begin{equation} 
M_i = |Q_i|  \ \ \ \ \ i=1,2 \label{critical}
\end{equation}
and balance can occur for arbitrary separation of the sources. Up to
the present time, the problem of gravitational-electrostatic balance
of two spherical bodies in general relativity without an intervening
Weyl line singularity (strut or tension) has been solved {\em exactly}
only for critically charged masses
\cite{coopcruz,papapetrou,tomimatsu}. A balance solution was
originally thought to have been found \cite{coopcarminati} within the
Herlt class for both sources having $M_i > \left|Q_i\right|$, but it
was subsequently shown that the intervening line singularity could not
be removed \cite{perry}. Kramer \cite{kramer} presented an exact
solution for the electrostatic counterpart of the double Kerr-NUT
solution with zero spin parameter.  He found that condition
(\ref{classical}) holds for electrostatic balance. However he stated
that his solution cannot be interpreted as the non-linear
superposition of two Reissner-Nordstr\"{o}m solutions and thus the
masses are not spherically symmetric.

In the present paper, an exact electrostatic solution of the
Einstein-Maxwell equations representing the exterior field of two
arbitrary charged non-linearly superposed Reissner-Nordstr\"{o}m
sources in equilibrium is given. It is obtained with the aid of
Sibgatullin's \cite{sib} method for constructing the complex Ernst
potentials \cite{ernst1}. It is mathematically equivalent to the
solutions of Manko {\em et~al} \cite{manko} and Chamorro {\em
et~al} \cite{chamorro}, henceforth referred to as papers~I and
papers~II respectively, (with their spin parameters set to zero) and
they are all special cases of the general mathematical solution given
by Ernst \cite{ernst2}. It is of primary importance that the parameters
in the solution be related to a {\em physical} set of parameters in
order for any subsequent analysis of the solution to have any
significant physical meaning. For a physical set of parameters, one
would prefer to use the individual masses and charges of each source
and the distance between the sources. The invariant charge enclosed by
a spacelike hypersurface can be found by the direct integration of
Maxwell's equations. For space-times with a timelike Killing vector, a
conserved quantity which can be interpreted as the contribution to the
total mass from each body can be invariantly defined in analogy with
the charge (See, for example, Refs.~\onlinecite{kramer,dietz}
and~\onlinecite{stephani}).  This paper follows Kramer \cite{kramer}
for the definition of the individual mass of each body. In
Section~\ref{massandcharge}, the integrals of charge and mass are
given and they are applied to the Weyl-class solution for two
Reissner-Nordstr\"{o}m bodies in Section~\ref{weylsection}.
Section~\ref{non-weyl} presents the solution for a parameterization of
the non-Weyl-class double Reissner-Nordstr\"{o}m solution based on the
Weyl-class parameterization. These are then compared to the
parameterizations proposed in papers~I and~II. It is shown that the
parameterizations employed in papers~I and~II do not represent the
physical masses or charges of the individual sources even in the
Weyl-class limit (except for the special case of identical bodies in
paper~I). Due to the complexities of the parameterization, a rendering
of the solution in terms of the individual masses and charges as given
in Section~\ref{massandcharge} has not yet been accomplished.
However, numerical analysis of the physical masses and charges is
possible for a given set of parameters.  In Section~\ref{equilibsec},
balance without a strut or tension for numerical values of the
physical mass and charge is examined. It is found that there are
balance conditions for which neither body is critically charged and
the Newtonian balance condition does not hold. This is in accordance with
Bonnor's \cite{bonnor1} test particle analysis. The dependence of the
balance condition on the separation of the bodies is not yet known. A
discussion of the results and conclusions are given in
Sections~\ref{discussion} and~\ref{conclusion}.

\section{Mass and Charge} \label{massandcharge}

For a static axially symmetric space-time, the mass $M_i$ and charge
$Q_i$ of a source inside a closed 2--surface $\sigma_i$ are given by
the integrals \cite{landau,kramer}
\begin{equation} \label{mass}
M_i \equiv -\frac{1}{8\pi} \oint_{\sigma_i} \!\! K^{ab} \sqrt{-g} \,
df^*_{ab}\, 
\end{equation}
\begin{equation} \label{charge}
Q_i = -\frac{1}{8\pi} \oint_{\sigma_i} \!\! F^{ab} \sqrt{-g} \,
df^*_{ab} 
\end{equation}
where 
\begin{equation}
K^{ab} \equiv \xi^{a;b}+\Phi F^{ab}\, .
\end{equation}
The timelike Killing vector is $\xi^a$, $F_{ab}$ is the
electromagnetic field tensor, $\Phi$ is the electrostatic potential,
$g$ is the determinant of the metric and $df^*_{ab}$ is the dual to
the surface element 2-form $df^{ab}$,
\begin{equation}
df^{*}_{ab} = \frac{1}{2} e_{abcd}df^{cd}
\end{equation}
(here $e_{abcd}$ is the flat space Levi--Civita permutation symbol).
The above integral conservation laws follow from the local
conservation laws 
\begin{equation}
F^{ab}_{\ \ ;b;a} = 0 \ \ \ \ \ \ K^{ab}_{\ \ ;b;a} = 0,
\end{equation}
the first, following from the conservation of charge and the second
from the existence of the timelike Killing vector $\xi^a$ and the
restriction to a static axially symmetric space-time metric. Since the
Einstein--Maxwell equations also imply
\begin{equation}
F^{ab}_{\ \ ;b} = 0 \ \ \ \ \ \ K^{ab}_{\ \ ;b} = 0,
\end{equation}
in a source free region, any deformation of the surface $\sigma_i$ in
the electrovacuum region outside the sources does not change the
values of the integrals $M_i$ and $Q_i$.

\section{The Weyl-Class Two Body Solution} \label{weylsection}

\begin{figure}
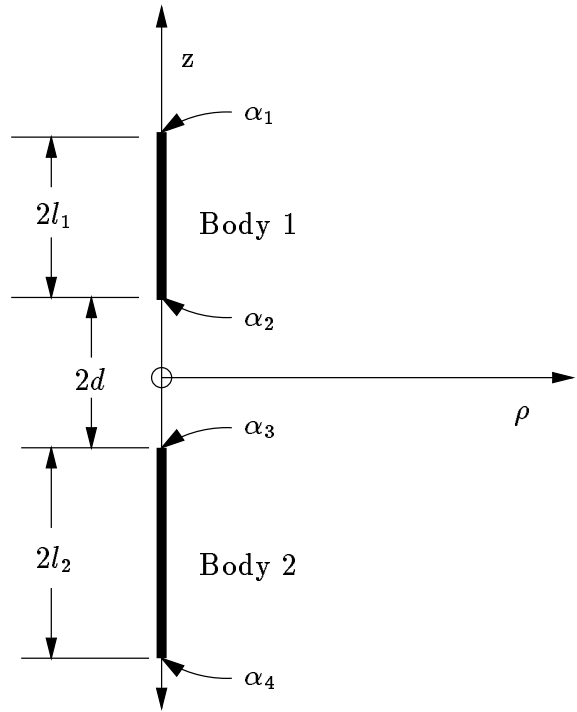

\caption{Schematic of two Reissner-Nordstr\"{o}m black holes in
cylindrical coordinates. The thick lines are the Weyl ``rods'' which
show the locations of the event horizon surfaces.}
\label{fig1}
\end{figure}

To investigate the structure of space-times with two sources, the
Weyl-class double Reissner-Nordstr\"{o}m solution provides a suitable
yet mathematically uncumbersome framework from which to proceed. The
solution is easily found through the method presented in
Ref.~\onlinecite{coopcruz}. The metric for a static axially symmetric
space-time can be written in the canonical form
\begin{equation} \label{metric}
ds^2 = e^w dt^2 - e^{v-w}\left(d\rho^2+dz^2\right) - \rho^2
e^{-w}d\phi^2, 
\end{equation}
where $w$ and $v$ are functions of the cylindrical coordinates $\rho$ and
$z$. The Weyl-class solutions are characterized by the metric
function $w$ being a function of the electrostatic potential, i.e. $w
= w(\Phi)$ so that the gravitational and electrostatic equipotential
surfaces overlap. For asymptotically flat boundary conditions,
the unique functional relationship between $e^w$ and $\Phi$
is \cite{weyl} 
\begin{equation} \label{weylclass}
e^w = 1 - 2\frac{m_{\mbox{\tiny T}}}{q_{\mbox{\tiny T}}} \Phi + \Phi^2,
\end{equation}
where $\Phi$ is the electrostatic potential and $m_{\mbox{\tiny T}}$
and $q_{\mbox{\tiny T}}$ are the total mass and charge
respectively. The solution representing two ``undercharged'' ($M_i >
\left| Q_i \right|$) Reissner-Nordstr\"{o}m bodies (or ``black
holes'') is given by
\begin{equation}
\Phi = a\frac{f-1}{a^2 f -1},
\end{equation}
where 
\begin{equation}
f = \left(\frac{R_1+R_2 - 2 l_1}{R_1+R_2 + 2 l_1}\right) 
    \left(\frac{R_3+R_4 - 2 l_2}{R_3+R_4 + 2 l_2}\right)\, ,
\end{equation}
\begin{eqnarray}
R^2_1 & \equiv & (z - d - 2 l_1)^2 + \rho^2 \, , \\
R^2_2 & \equiv & (z - d )^2 + \rho^2 \, , \\
R^2_3 & \equiv & (z + d )^2 + \rho^2 \, , \\
R^2_4 & \equiv & (z + d + 2 l_2)^2 + \rho^2 \, . 
\end{eqnarray}
The constant parameters $2d$ and $2l_1,2l_2$ are the coordinate
distance between the horizons and the ``lengths'' of the horizons
(Weyl ``rods'') respectively (see Fig.~\ref{fig1}). The parameter $a$
is defined through the equation
\begin{equation}
\frac{1+a^2}{a} = \frac{2 m_{\mbox{\tiny T}}}{q_{\mbox{\tiny T}}} .
\end{equation}
The metric function $e^w$ is found through Eq.~(\ref{weylclass}).
The metric function $e^v$ is
\[
e^v = \frac{\left(R_1+R_2\right)^2-4l_1^2}{4 R_1 R_2} \cdot
\frac{\left(R_3+R_4\right)^2-4l_2^2}{4 R_3 R_4} \cdot 
\left[ \frac{\left(\left(l_1+l_2+d\right) R_1 + \left(l_2+d\right) R_2
- l_1 R_4\right)d}{\left(\left(l_1+d\right) R_1 + R_2 d -l_1 R_3
\right) \left(l_2 + d\right)}\right]^2 \; . 
\]
Choosing the surface $\sigma_1$ to encompass body 1 and the surface
$\sigma_2$ to encompass body 2 of Fig.~\ref{fig1}, the mass and charge
integrals of Eqs.~(\ref{mass}) and (\ref{charge}) yield
\begin{equation}
\begin{array}{rclrcl}
M_1 & = &{\displaystyle \frac{1+a^2}{1-a^2} }\, l_1 & \mbox{\ \ \ }
Q_1 & = &{\displaystyle \frac{2a}{1-a^2} }\, l_1 \\ \mbox{ } \\
M_2 & = &{\displaystyle \frac{1+a^2}{1-a^2} }\, l_2 &  \mbox{\ \ \ }
Q_2 & = &{\displaystyle \frac{2a}{1-a^2} }\, l_2
\end{array}
\end{equation}
The above form of the individual mass and charge for each
Reissner-Nordstr\"{o}m body is similar to the form proposed in
Ref.~\onlinecite{coopcruz} for the mass and charge decomposition of
two charged Curzon particles. It was stated in
Ref.~\onlinecite{coopcarminati} that the conjectured charge
decomposition for both the double Reissner-Nordstr\"{o}m and Curzon
cases were verified by direct calculation through
Eq.~(\ref{charge}). It is straightforward to verify that
Eq.~(\ref{mass}) yields the conjectured mass decomposition for the
double Curzon solution. Because of the functional relationship between
the gravitational potential and the electrostatic potential, not all
of the parameters $M_1,M_2,Q_1,Q_2$ are independent. Thus the
Weyl-class is also characterized by the constraint
\begin{equation} \label{weylconstraint}
M_1 Q_2 = M_2 Q_1.
\end{equation}
Removal of the line singularity between the bodies yields
Eq.~(\ref{classical}) as an additional condition on the parameters. As
a result, the parameters also satisfy Eq.~(\ref{critical}). Thus
equilibrium without a strut or tension occurs for ``critically''
charged sources and this balance is found to be independent of the
separation distance \cite{coopcruz}.

\section{Non-Weyl parameterizations} \label{non-weyl}

Generalizing the Weyl-class double Reissner-Nordstr\"{o}m solution to
the case in which the gravitational and electrostatic equipotential
surfaces no longer overlap has usually been attempted through the
means of generating techniques (see, for example,
Refs.~\onlinecite{tomimatsu,coopcarminati} and
\onlinecite{kramer}). In these techniques, new solutions are generated
from old ones rather than by solving the equations directly. Recently,
considerable interest has focused upon a method \cite{sib} which
constructs the Ernst potentials \cite{ernst1} from initial data on the
symmetry axis. The complex Ernst potentials ${\cal E}(\rho,z) $ and
$\Psi(\rho,z)$ of all stationary axisymmetric electrovacuum
space-times with axis data of the form
\begin{equation}
{\cal E}(z,\rho=0) = \frac{U-W}{U+W}, \ \ \ \Psi (z,\rho=0) =
\frac{V}{U+W} ,
\end{equation}
where
\begin{equation}
U = z^2 + U_1 z + U_2 
\end{equation}
\begin{equation}
V = V_1 z + V_2 
\end{equation}
\begin{equation}
W = W_1 z + W_2 
\end{equation}
and $U_1,U_2,V_1,V_2,W_1,W_2$ are complex constants, have been
found \cite{ernst2}. However, a mathematical solution to the
Einstein-Maxwell field equations does not imply a well understood
physical interpretation of the solution. Sibgatullin's method of
constructing the Ernst potentials aids in obtaining the physically
meaningful parameterization which is sought for the two-body case in
question.

In Sibgatullin's method, it is required that the Ernst potentials along
the $z$-axis be specified. Our choice was \cite{perrycoop1,perrycoop2} 
\[
{\cal E}(\rho=0,z) \equiv e(z) = 
1-\frac{2(m_1(z+z_2) + m_2(z+z_1))}{(z+z_1+m_1)(z+z_2+m_2) -q_1 q_2},
\]
\begin{equation} \label{ourernst}
\Psi (\rho=0,z) \equiv F(z) = 
\frac{q_1(z+z_2) + q_2(z+z_1)}{(z+z_1+m_1)(z+z_2+m_2) -q_1 q_2}.
\end{equation}
It has the form of the Weyl-class double Reissner-Nordstr\"{o}m axis
data. If the additional Weyl-class constraint
\begin{equation} \label{weylcondition}
m_1 q_2 - m_2 q_1 = 0
\end{equation}
is placed on the functions $e(z)$ and $F(z)$, then Sibgatullin's
method yields the Weyl-class double Reissner-Nordstr\"{o}m solution
(in an alternate form to Ref.~\onlinecite{coopcruz}) and the
parameters $m_1,m_2,q_1,q_2$ are the physical masses and charges as
defined by Eqs.~(\ref{mass}--\ref{charge}) (i.e.,
$M_1=m_1,Q_1=q_1,M_2=m_2,Q_2=q_2$). For the solution of two Weyl-class
Reissner-Nordstr\"{o}m black holes (given in
Section~\ref{weylsection}), Fig.~\ref{fig1} shows the coordinate
positions of the centers of the ``rods'' as $d+l_1$ for body~1 and
$-d-l_2$ for body~2.  The parameters $z_1,z_2$ identify the negative
of the coordinate positions of the centers of the ``rods'', i.e.,
\begin{eqnarray*}
z_1 & = & -d-l_1, \\
z_2 & = & d + l_2.
\end{eqnarray*}
If condition~(\ref{weylcondition}) is not imposed, $w\neq w(\Phi)$,
i.e., the gravitational and electrostatic equipotential surfaces no
longer overlap. In Section~\ref{equilibsec} it will be shown that the
parameters $m_1,m_2,q_1,q_2$ then no longer carry the suggested
physical meaning and the parameters $z_1,z_2$ no longer coincide with
the centers of the ``rods'' when the Weyl-class
constraint~(\ref{weylcondition}) is not imposed.

The full Ernst potentials ${\cal E}(\rho,z)$ and $\Psi(\rho ,z)$ for
the axis data of Eq.~(\ref{ourernst}), expressed in terms of the
cylindrical coordinates $(\rho,\,z)$, are found to be (the details of
the method can be found in Refs.~\onlinecite{sib,manko,mankosib} and
in the review article \onlinecite{mankosibreview})
\begin{equation} \label{ernstpot}
{\cal E} = \frac{A-B}{A+B}, \ \ \ \Psi = \frac{C}{A+B},   
\end{equation}
where
\[
A \equiv \sum_{i < j}^4 a_{ij} r_i r_j, \ \ 
B \equiv \sum_{i =1}^4 b_i r_i , \ \ 
C \equiv \sum_{i =1}^4 c_i r_i ,
\ \ r_n \equiv \sqrt{\rho^2 + (z-\alpha_n)^2} ,\ \ \ (n = 1
\rightarrow 4).
\]
The constants
$\alpha_n$ in Eq.~(\ref{ernstpot}) are the roots of the equation
\begin{equation}
e(z)+\left[F(z)\right]^2=0
\end{equation}
and can only be real or complex conjugate pairs.  The remaining
constants $a_{ij},\, b_i$ and $c_i$ are defined as follows:
\[
\begin{array}{l} a_{ij} \equiv (-1)^{i+j+1} s_i s_j t_i t_j 
(s_i t_j - s_j t_i)
\left|\begin{array}{cc} s_k v_k & s_l v_l \\ t_k u_k & t_l u_l
\end{array}\right|, \\
\mbox{} \\
(i<j; \ k < l; \ k,l\neq i,j; \ \;i,k=1\rightarrow
3;\;j,l=2\rightarrow 4); \\ 
\mbox{} \\
b_i \equiv (-1)^{i} s_i t_i (s_i - t_i)
\left|\begin{array}{ccc} s^2_k t^2_k & s^2_l t^2_l & s^2_m t^2_m \\
 s_k v_k & s_l v_l & s_m v_m \\ t_k u_k & t_l u_l & t_m u_m
\end{array}\right|, \\
\mbox{} \\
(k<l<m; \ k,l,m \neq i;\ \;i=1\rightarrow 4;\;k=1,2; \;l=2,3;\;m=3,4);
\\ 
\mbox{} \\
c_i \equiv (-1)^{i+1} s_i t_i (s_i - t_i)(K_3 G_i + K_4 H_i), 
\end{array}
\]
\begin{equation} \label{ernstpotconst}
G_i \equiv \left|\begin{array}{ccc} s_k t^2_k & s_l t^2_l & s_m t^2_m \\
 s_k v_k & s_l v_l & s_m v_m \\ t_k u_k & t_l u_l & t_m u_m
\end{array}\right|, 
\ \ \ \ 
H_i \equiv \left|\begin{array}{ccc} s^2_k t_k & s^2_l t_l & s^2_m t_m \\
 s_k v_k & s_l v_l & s_m v_m \\ t_k u_k & t_l u_l & t_m u_m
\end{array}\right|, 
\end{equation}
\[
(k<l<m; \ k,l,m \neq i;\ \;i=1\rightarrow 4;\;k=1,2; \;l=2,3;\;m=3,4);
\\ 
\]
\[
s_i \equiv \beta_1-\alpha_i, \ \ t_i \equiv \beta_2-\alpha_i, 
\]
\[
u_i \equiv K_1 s_i t_i + K_3^2 t_i + K_3 K_4 s_i, \ \ 
v_i \equiv K_2 s_i t_i + K_4^2 s_i + K_3 K_4 t_i ,
\]
\[
K_1 \equiv \frac{m_1 z_2+m_2 z_1 + (m_1 +m_2) \beta_1}{\beta_1-\beta_2}
,
\ \ 
K_2 \equiv \frac{m_1 z_2+m_2 z_1 + (m_1 +m_2) \beta_2}{\beta_2-\beta_1},
\]
\[
K_3 \equiv \frac{q_1 z_2+q_2 z_1 + (q_1 +q_2) \beta_1}{\beta_1-\beta_2}
,
\ \ 
K_4 \equiv \frac{q_1 z_2+q_2 z_1 + (q_1 +q_2) \beta_2}{\beta_2-\beta_1},
\]
\[
\beta_1 \equiv -\frac{1}{2} \left( z_1 + m_1+z_2+m_2 - \sqrt{ \left(
z_1-z_2 +m_1-m_2\right)^2 + 4 q_1 q_2} \right),
\]
\[
\beta_2 \equiv -\frac{1}{2} \left( z_1 + m_1+z_2+m_2 + \sqrt{ \left(
z_1-z_2 +m_1-m_2\right)^2 + 4 q_1 q_2} \right),
\]
where all of the subsequent quantities introduced are constants
ultimately defined in terms of $m_i$, $q_i$, $z_i$, $i=1,2$, which
specify the character and locations of the sources in the Weyl-class
limit only.


The expressions for $\cal E$ and $\Psi$ are in Kinnersley's
\cite{kinnersley} form and this permits one to write the
corresponding metric functions as
\begin{equation}
e^w = \frac{A \bar{A} - B\bar{B}+C\bar{C}}{(A+B)(\bar{A}+\bar{B})}, 
\ \ 
e^v = \frac{A \bar{A} - B\bar{B}+C\bar{C}}{K_0 r_1 r_2 r_3 r_4},
\end{equation}
where
\begin{equation}
K_0 = \left(\sum^4_{i<j} a_{ij}\right)\left(\sum^4_{i<j}
\bar{a}_{ij}\right) 
\end{equation}
and a bar denotes complex conjugation. For a static metric, the
electrostatic potential $\Phi$ is equal to the Ernst potential $\Psi$
and this completes the solution.

With the knowledge of the full Ernst potentials and the metric
functions, the next step would be to evaluate the true mass and charge
integrals in terms of the parameters $m_1,m_2,q_1,q_2,z_1,z_2$. It is
to be stressed that outside of the Weyl-class, these parameters no
longer carry the suggested physical meaning.  For the metric
(\ref{metric}), the integrals (\ref{mass}) and (\ref{charge}) can be
written as relations in flat 3-space ($i=1,2$):
\begin{equation} \label{flatmass}
M_i = \frac{1}{8\pi} \oint_{\sigma_i} \!\! e^{-w} {\cal E}_{,\alpha}
\, n^{\alpha} \, dA 
\end{equation}
\begin{equation} \label{flatcharge}
Q_i = -\frac{1}{4\pi} \oint_{\sigma_i} \!\! e^{-w} {\Phi}_{,\alpha}
\, n^{\alpha} \, dA  \, ,
\end{equation}
where $n^{\alpha}$ ($\alpha$ runs from 1 to 3) is the unit vector
orthogonal to the surface and $dA$ denotes the invariant (flat)
surface element (see also Ref.~\onlinecite{kramer} and references
therein). 

\begin{figure}
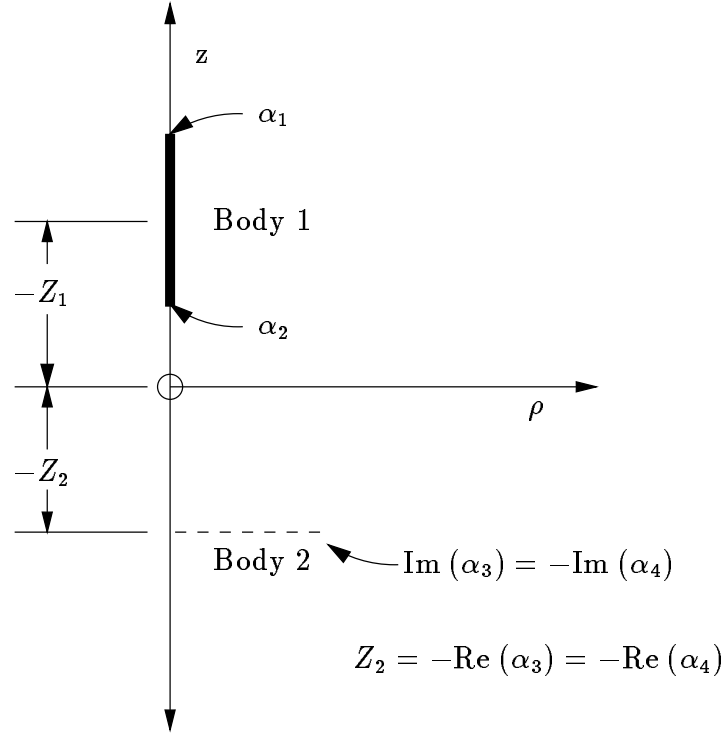

\caption{Schematic of a Reissner-Nordstr\"{o}m black hole and a
Reissner-Nordstr\"{o}m superextreme body. The dotted line is a
``complex Weyl rod''. The intersection of the ``rod'' with the
$z$-axis is defined as the coordinate position of body 2.}
\label{fig2}
\end{figure}

We can extend the Weyl-class definitions of the coordinate positions
of the bodies to the non-Weyl-class solution. There are three distinct
types of sources of interest. They are characterized by the transition
between a source with an event horizon to one without an event
horizon. As mentioned previously, the constants $\alpha_n,
n=1\rightarrow 4$ in Eq.~(\ref{ernstpot}) are either real or complex
conjugate pairs. By definition, we choose $\alpha_1 \geq \alpha_2 >
\alpha_3 \ge \alpha_4$. A Reissner-Nordstr\"{o}m ``black hole'' is
characterized by real pairs of $\alpha_n$. Fig.~\ref{fig1} shows that,
in the Weyl canonical coordinate system, $\alpha_n$ indicates the end
points of a ``Weyl rod'', which itself is the event horizon surface. A
``superextreme'' object \cite{manko} or ``naked singularity'' is
characterized by a complex conjugate pair of $\alpha_n$. Body~2 of
Fig.~\ref{fig2} illustrates the manifestation of a ``superextreme''
body in the space-time.  An ``extreme'' object, for example, would be
characterized by real $\alpha_n$ for which $\alpha_1 = \alpha_2$.
Therefore we have the following definitions for the coordinate
positions of the sources:
\begin{description}
\item[{\it i})] For a Reissner-Nordstr\"{o}m ``black hole'', we define
   $-Z_i$ to be the coordinate position of the center of the ``Weyl
   rod''. For example, the coordinate position of body ~1 of
   Fig.~(\ref{fig2}) is 
   \[ 
   -Z_1 = \frac{1}{2}\left(\alpha_1+\alpha_2\right).
   \]

\item[{\it ii})] For a ``superextreme'' object, we define $-Z_i$ to be
   the coordinate position of the real part of $\alpha_n$. For
   example, body~2 of Fig.~(\ref{fig2}) is a ``superextreme''
   object. Therefore its coordinate position is
   \[
   -Z_2 = \mbox{Re}(\alpha_3) = \mbox{Re}(\alpha_4).
   \]
   
   (One could consider the imaginary part of $\alpha_n$ as the end
   points of a ``complex Weyl rod'' with the coordinate position of
   this ``complex rod'' being defined as its intersection with the
   real axis ($z$-axis).)

\item[{\it iii})] For an ``extreme object'', we define $-Z_i$ to be
   the coordinate position of the point locating the zero ``length''
   Weyl ``rod''. For example, if body~1 was an ``extreme'' object,
   then $\alpha_1 = \alpha_2$ and $-Z_1 = \alpha_1$.

\end{description}
We also define
\begin{equation} 
\mbox{Re}(\alpha_2) > \mbox{Re}(\alpha_3) 
\end{equation} 
as the condition for having two separated bodies irrespective of the
type of object.

With the above integrals and the coordinate positions as defined above
evaluated in terms of $m_1,m_2,q_1,q_2,z_1,z_2$, it would then be
possible, in principle, to invert these equations and hence write the
solution (\ref{ernstpot}--\ref{ernstpotconst}) in terms of the true
physical parameters $M_i,Q_i$ and the coordinate positions $Z_i,\;
i=1,2$. Ideally, the coordinate positions of the sources should be
replaced with the proper separation of the sources. The complexity of
the above Ernst potentials makes the analytic evaluation of the
integrals (\ref{flatmass}-\ref{flatcharge}) and the proper separation
difficult. As a consequence this goal has not yet been
achieved. However, it is possible to numerically integrate
Eqs.~(\ref{mass}--\ref{charge}) for a given set
$\{m_1,m_2,q_1,q_2,z_1,z_2\}$. This will prove to be useful in
studying balance conditions without a strut in
Section~\ref{equilibsec}.

Although the numerical evaluation of the physical mass and charge can
be achieved from the parameterizations of paper~I or paper~II, it was
hoped that the parameterization proposed in this paper, based on the
Weyl-class solution, would facilitate the analytic evaluation of the
integrals. It is not difficult to show that the parameterizations in
papers~I or II do not correctly identify the individual masses and
charges of each source. We stated earlier that our parameterization
$\{m_1,m_2,q_1,q_2,z_1,z_2\}$ only represents the physical masses and
charges and coordinate positions of each source when the Weyl-class
condition (Eq.~(\ref{weylclass}) or (\ref{weylcondition})) is imposed
(i.e., $\{M_1=m_1,M_2=m_2,Q_1=q_1,Q_2=q_2,Z_1=z_1,Z_2=z_2\}$) . We can
best demonstrate the problems with the parameterizations of papers~I
and II by comparing the representation of a properly parameterized
Weyl-class solution with each of the other parameterizations. Let the
set $\{m_1,m_2,q_1,q_2,z_1,z_2\}$ represent the physical Weyl-class
parameters under the condition $m_1 q_2 = m_2 q_1$. Then the
relationships between the three parameterizations is found by solving
the set of equations (setting the spin parameters found in papers~I
and II to zero)
\begin{equation} \label{eqnsystem}
\begin{tabular}{ccccc}
\multicolumn{1}{c}{\underline{Weyl-class}} & &
\multicolumn{1}{c}{\underline{Paper~I}} & 
& \multicolumn{1}{c}{\underline{Paper~II}} \\ 
$m_1 + m_2$ & = & $\tilde{m}_1 + \tilde{m}_2$ & = & $\hat{m}_1 + \hat{m}_2 $\\ 
$q_1 + q_2$ & = & $\tilde{q}_1 + \tilde{q}_2$ & = & $\hat{q}_1 + \hat{q}_2 $\\ 
$z_1 + z_2$ & = & $\tilde{z}_1 + \tilde{z}_2$ & = & $\hat{z}_1 + \hat{z}_2 $\\ 
$m_1 z_2 + m_2 z_1$ & = & $\tilde{m}_1\tilde{z}_2 +
\tilde{m}_2\tilde{z}_1$ & = & $\hat{m}_1\hat{z}_2 + \hat{m}_2\hat{z}_1 +
2 \hat{m}_1\hat{m}_2 $\\ 
$q_1 z_2 + q_2 z_1 $& = & $\tilde{q}_1\tilde{z}_2 +
\tilde{q}_2\tilde{z}_1$ & = & 
$\hat{q}_1\hat{z}_2 + \hat{q}_2\hat{z}_1  + \hat{q}_1\hat{m}_2 +
\hat{q}_2\hat{m}_1$  \\
$z_1 z_2 + m_1 m_2 - q_1 q_2$ & = & $\tilde{z}_1 \tilde{z}_2 +
\tilde{m}_1 \tilde{m}_2$ & = &
$\hat{z}_1 \hat{z}_2 - \hat{m}_1\hat{m}_2 $ .
\end{tabular}
\end{equation}
The tilded and careted parameters are the parameterizations of
papers~I and~II respectively. Table~\ref{table1} summarizes the
results of solving the system (\ref{eqnsystem}) given the values shown
in column~1. The solution represents two Weyl-class
Reissner-Nordstr\"{o}m ``critically charged'' bodies without an
intervening line singularity. It is clear that none of the parameter
values in the latter two columns match the physical Weyl-class
values. In fact one has to assign negative values to $\tilde{m}_2,
\tilde{q}_2$ in order to obtain a {\em positive} physical mass and
charge for source 2.  Thus, apart from one special case, neither the
paper~I nor the paper~II parameterizations can be interpreted in as
the invariant physical parameters. The only exception is for identical
bodies (with or without a line singularity) in the parameterization of
paper~I. In this very special case of the Weyl-class, the parameters
$\tilde{m}_1 =\tilde{m}_2,\;\tilde{q}_1 =\tilde{q}_2$ are the physical
masses and charges. However, $\tilde{z}_1$ and $\tilde{z}_2$ do not
identify the coordinate positions of the bodies as defined
earlier. The paper~II parameterization is not physical even for
identical bodies.

It is the demand for the inclusion of the Weyl-class solution in
Ref.~\onlinecite{coopcruz} which led to our form of $e(z)$ and $F(z)$.
It should be emphasized that our parameterization contains as a
special case, the simplest clearly individually spherical two-body
balance solution of two critically charged bodies. This can be best
illustrated by examining the Simon \cite{simon,hoen} relativistic
multipole moments of each parameterization.  The first five Simon
relativistic mass and charge multipole moments for our
parameterization are
\begin{equation}   \label{massmultipoles}
\begin{array}{ccl}
{\cal M}_0 & = & m_1+m_2, \\ 
{\cal M}_1 & = & \mbox{} - m_1 z_1 - m_2 z_2, \\ 
{\cal M}_2 & = & m_1 z_1^2 + m_2 z_2^2 - \left(m_1 m_2 - q_1 q_2\right)
\left(m_1 + m_2\right), \\ 
{\cal M}_3 & = & \mbox{} - m_1 z_1^3 - m_2 z_2^3 + \left(m_1 m_2 - q_1
q_2\right) 
\left(2 m_1 z_1 + 2 m_2 z_2 + z_1 m_2 + z_2 m_1 \right), \\ 
{\cal M}_4 & = & m_1 z_1^4 + m_2 z_2^4 - \left(m_1 m_2 - q_1 q_2\right) 
\Big[ \left(m_1 +m_2 \right) \left(q_1 q_2 - m_1 m_2\right) \\ 
\mbox{} & \mbox{} & \mbox{}
+ 2 \left(m_1 z_1^2 + m_2 z_2^2\right) 
+ \left(m_1+m_2\right)
\left(z_1+z_2\right)^2 \\ 
\mbox{} & \mbox{} & \mbox{}
+ \left. \frac{1}{7} \left(m_1+m_2\right)\left(\left(q_1+q_2\right)^2-
\left(m_1+m_2\right)^2\right)\right] \\ 
\mbox{} & \mbox{} & \mbox{}
-\frac{1}{210}\left(z_1-z_2\right)\left[16\left(z_1-z_2\right)
\left(m_1+m_2\right)\left(m_1
q_2 - m_2 q_1\right)^2 \right.
\\  \mbox{} & \mbox{} & \mbox{}
+ z_1\left(30 m_1 \left(m_1 m_2 + m_2^2 - q_2^2\right) -3 q_1\left( 3
m_2 q_1 + 7 q_2 m_1\right)\right) \\ 
\mbox{} & \mbox{} & \mbox{}
-  z_2\left(30 m_2 \left(m_1 m_2 + m_1^2 - q_1^2\right) -3 q_2\left( 3
m_1 q_2 + 7 q_1 m_2\right)\right) \Big]
\end{array}
\end{equation}
and
\begin{equation}   \label{chargemultipoles}
\begin{array}{ccl}
{\cal Q}_0 & = & q_1+q_2,  \\ 
{\cal Q}_1 & = & \mbox{} - q_1 z_1 - q_2 z_2 ,   \\ 
{\cal Q}_2 & = & q_1 z_1^2 + q_2 z_2^2 - \left(m_1 m_2 - q_1 q_2\right)
\left(q_1 + q_2\right),   \\ 
{\cal Q}_3 & = & \mbox{} - q_1 z_1^3 - q_2 z_2^3 + \left(m_1 m_2 - q_1
q_2\right) 
\left(2 q_1 z_1 + 2 q_2 z_2 + z_1 q_2 + z_2 q_1 \right).   \\ 
{\cal Q}_4 & = & q_1 z_1^4 + q_2 z_2^4 - \left(m_1 m_2 - q_1 q_2\right) 
\Big[ \left(q_1 +q_2 \right) \left(q_1 q_2 - m_1 m_2\right)   \\ 
\mbox{} & \mbox{} & \mbox{}
+ 2 \left(q_1 z_1^2 + q_2 z_2^2\right) + \left(q_1+q_2\right)
\left(z_1+z_2\right)^2   \\ 
\mbox{} & \mbox{} & \mbox{}
+ \frac{1}{7} \left. \left(q_1+q_2\right)\left(\left(q_1+q_2\right)^2-
\left(m_1+m_2\right)^2\right)\right]   \\ 
\mbox{} & \mbox{} & \mbox{}
-\frac{1}{210}\left(z_1-z_2\right)\left[16\left(z_1-z_2\right)
\left(q_1+q_2\right)\left(m_1
q_2 - m_2 q_1\right)^2 \right.  
\\ \mbox{} & \mbox{} & \mbox{}
- z_1\left(30 q_2 \left(q_1 q_2 - m_1 m_2 + q_1^2\right) -3 m_1\left( 13
m_1 q_2 - 3 m_2 q_1\right)\right)   \\ 
\mbox{} & \mbox{} & \mbox{}
+ z_2\left(30 q_1 \left(q_1 q_2 - m_1 m_2 + q_1^2\right) -3 m_2\left( 13
m_2 q_1 - 3 m_1 q_2\right)\right)\Big] 
\end{array} 
\end{equation}
respectively. In Newtonian physics, a system of two monopoles at
positions $z_1,\ z_2$ has multipole moments
\begin{equation} \label{newtpoles} 
{\cal M}_n = m_1 z_1^n + m_2 z_2^n, \ \ \ \ {\cal Q}_n = q_1 z_1^n +
q_2 z_2^n.   
\end{equation}
It is interesting to observe that this is also the relativistic
multipole structure for two Weyl-class critically charged bodies, at
least up to ${\cal M}_4,\ {\cal Q} _4$. There is an inherent
asphericity imposed upon each, since the two bodies are interacting in
a line.  For non-linearly interacting sources in a line, one would not
expect to realize perfect sphericity of the individual sources. (It is
yet to be explained why the sphericity is maintained in the
Weyl-class, at least up to ${\cal M}_4,\ {\cal Q} _4$.) Once the
solution is written analytically in terms of the physically meaningful
constants $M_i,Q_i$ and the coordinate positions $Z_i,\; i=1,2$,
one will be able to examine the general multipole structure of
non-linearly interacting spherical bodies.

For comparison, the first four Simon relativistic mass and charge
multipole moments for the parameterization of paper~I (with their spin
parameters $a_i = 0, i=1,2$) are
\begin{equation}   \label{mankomassmultipoles}
\begin{array}{ccl}
{\cal M}_0 & = & \tilde{m}_1+\tilde{m}_2, \\ 
{\cal M}_1 & = & \mbox{} - \tilde{m}_1 \tilde{z}_1 - \tilde{m}_2
\tilde{z}_2, \\  
{\cal M}_2 & = & \tilde{m}_1 \tilde{z}_1^2 + \tilde{m}_2 \tilde{z}_2^2
- \tilde{m}_1 \tilde{m}_2  
\left(\tilde{m}_1 + \tilde{m}_2\right), \\ 
{\cal M}_3 & = & \mbox{} - \tilde{m}_1 \tilde{z}_1^3 - \tilde{m}_2
\tilde{z}_2^3 + \tilde{m}_1 \tilde{m}_2  
\left(2 \tilde{m}_1 \tilde{z}_1 + 2 \tilde{m}_2 \tilde{z}_2 +
\tilde{z}_1 \tilde{m}_2 + \tilde{z}_2 \tilde{m}_1 \right) 
\end{array}
\end{equation}
and
\begin{equation}   \label{mankochargemultipoles}
\begin{array}{ccl}
{\cal Q}_0 & = & \tilde{q}_1+\tilde{q}_2,  \\
{\cal Q}_1 & = & \mbox{} - \tilde{q}_1 \tilde{z}_1 - \tilde{q}_2
\tilde{z}_2 ,   \\  
{\cal Q}_2 & = & \tilde{q}_1 \tilde{z}_1^2 + \tilde{q}_2 \tilde{z}_2^2
- \tilde{m}_1 \tilde{m}_2  
\left(\tilde{q}_1 + \tilde{q}_2\right),   \\ 
{\cal Q}_3 & = & \mbox{} - \tilde{q}_1 \tilde{z}_1^3 - \tilde{q}_2
\tilde{z}_2^3 + \tilde{m}_1 \tilde{m}_2  
\left(2 \tilde{q}_1 \tilde{z}_1 + 2 \tilde{q}_2 \tilde{z}_2 +
\tilde{z}_1 \tilde{q}_2 + \tilde{z}_2 \tilde{q}_1 \right).  
\end{array} 
\end{equation}
The first four Simon relativistic mass and charge multipole moments
for the parameterization of paper~II (with their spin parameters
$a_i = 0, i=1,2$) are
\begin{equation}   \label{chamorromassmultipoles}
\begin{array}{ccl}
{\cal M}_0 & = & \hat{m}_1+\hat{m}_2, \\ 
{\cal M}_1 & = & \mbox{} - \hat{m}_1 \hat{z}_1 - \hat{m}_2 \hat{z}_2 +
2 \hat{m}_1 \hat{m}_2, \\  
{\cal M}_2 & = & \hat{m}_1 \hat{z}_1^2 + \hat{m}_2 \hat{z}_2^2 +
\hat{m}_1 \hat{m}_2 
\left(\hat{m}_1 + \hat{m}_2 -2 \hat{z}_1 -2 \hat{z}_2\right), \\  
{\cal M}_3 & = & \mbox{} - \hat{m}_1 \hat{z}_1^3 - \hat{m}_2
\hat{z}_2^3 \\ 
\mbox{} & \mbox{} & \mbox{} + \hat{m}_1 \hat{m}_2 
\left(2 \hat{m}_1 \hat{m}_2 + 2 \hat{z}_1 \hat{z}_2 + 2 \hat{z}_1^2 +2
\hat{z}_2^2 - \hat{m}_1 \hat{z}_2 - \hat{m}_2 \hat{z}_1 - 2 
\hat{m}_1 \hat{z}_1 - 2 \hat{m}_2 \hat{z}_2 \right) 
\end{array}
\end{equation}
and
\begin{equation}   \label{chammorochargemultipoles}
\begin{array}{ccl}
{\cal Q}_0 & = & \hat{q}_1+\hat{q}_2,  \\
{\cal Q}_1 & = & \mbox{} - \hat{q}_1 \hat{z}_1 - \hat{q}_2 \hat{z}_2
+\hat{m}_1 \hat{q}_2 + \hat{m}_2 \hat{q}_1, \\  
{\cal Q}_2 & = & \hat{q}_1 \hat{z}_1^2 + \hat{q}_2 \hat{z}_2^2 +
\hat{m}_1 \hat{m}_2 \left(\hat{q}_1 + 
\hat{q}_2\right) -\left(\hat{q}_1 \hat{m}_2 + \hat{q}_2
\hat{m}_1\right)\left(\hat{z}_1 + \hat{z}_2\right), \\  
{\cal Q}_3 & = & \mbox{} - \hat{q}_1 \hat{z}_1^3 - \hat{q}_2
\hat{z}_2^3 - \hat{m}_1 \hat{m}_2  
\left(2 \hat{q}_1 \hat{z}_1 + 2 \hat{q}_2 \hat{z}_2 + \hat{z}_1
\hat{q}_2 + \hat{z}_2 \hat{q}_1 \right) \\ 
\mbox{} & \mbox{} & \mbox{} + \left(\hat{q}_1 \hat{m}_2 + \hat{q}_2
\hat{m}_1\right) \left(\hat{m}_1 
\hat{m}_2 + \hat{z}_1 \hat{z}_2 + \hat{z}_1^2 +\hat{z}_2^2\right) . 
\end{array} 
\end{equation}
If the above parameterizations did represent the physical mass and
charge, it is evident that the multipole structure would not be that
of Newtonian spherical bodies even for critically charged bodies. As
stated earlier, it should be noted that in the parameterization of
paper~I, it can be shown that only in the case of identical bodies,
the parameters $\tilde{m}_1=\tilde{m}_2, \tilde{q}_1=\tilde{q}_2$ are
the physical mass and charge. However, in this case the multipoles
still do not have the form of Eq.~(\ref{newtpoles}) since the
parameters $\tilde{z}_1$ and $\tilde{z}_2$ do not identify the
positions of the bodies as defined earlier. A simple transformation
would correct the multipoles in this case.

\section{The Equilibrium Condition} \label{equilibsec}

In order for the space-time to be regular on the $z$-axis between the
sources (removal of the Weyl line singularity or imposition of the
condition for elementary flatness \cite{synge}), it is required that
the metric function
\begin{equation} \label{bc1}
v(z,\rho =0) = 0 
\end{equation}
between the sources.  If the origin of the coordinate system is
located between the sources (i.e., Re($\alpha_2$) $> 0$,
Re($\alpha_3$) $< 0$), then application of Eq.~(\ref{bc1}), after some
simplification, yields the balance equation
\begin{equation} \label{K6eqn}
K \equiv \frac{a_{12}\left(\bar{a}_{13}+\bar{a}_{14}\right)+\bar{a}_{12}
\left(a_{13}+a_{14}\right)}{K_0} = 0.
\end{equation}
Three cases were examined: $i$) Two Reissner-Nordstr\"{o}m black
holes, $ii$) Two Reissner-Nordstr\"{o}m superextreme bodies and $iii$)
One black hole and one superextreme body.

The procedure for testing for equilibrium without an intervening strut
or tension will be as follows:
\begin{enumerate}
\item Assign numerical values to five of the six parameters from the
unphysical set $\{m_1,m_2,q_1,q_2,z_1,z_2\}$.
\item Solve Eq.~(\ref{K6eqn}) for the unknown variable. 
\item If a real root of Eq.~(\ref{K6eqn}) exists, then evaluate
Eqs.~(\ref{mass}) and (\ref{charge}) to determine the physical mass
and charge parameters.
\end{enumerate}
The results for each of the three cases are as follows:

\subsection{Two Reissner-Nordstr\"{o}m Black Holes}

Numerous sets of the parameters $\{m_1,m_2,q_1,q_2,z_1,z_2\}$, such
that the constants $\alpha_n, n=1\rightarrow 4$ are real, were
investigated. No roots were found of Eq.~(\ref{K6eqn}). For example,
choosing $m_1=9.0,\;\; q_1=3.0,\;\; z_1=-15.0, \;\; m_2=8.0, \;\;
q_2=2.0$, no balance for $0\leq z_2 \leq 10^{10}$ was found. These
findings are consistent with other results \cite{tomimatsu,perry,dietz}
that two Reissner-Nordstr\"{o}m black holes cannot be found in
equilibrium without an intervening strut or tension.

\subsection{Two Reissner-Nordstr\"{o}m Superextreme Bodies}

Numerous sets of the parameters $\{m_1,m_2,q_1,q_2,z_1,z_2\}$, such
that the constants $\alpha_n, n=1\rightarrow 4$ are complex conjugate
pairs, were investigated. No roots were found of
Eq.~(\ref{K6eqn}). For example, in choosing $m_1=3.0,\;\; q_1=9.0,\;\;
z_1=-15.0, \;\; m_2=2.0, \;\; q_2=8.0$, no balance for $0\leq
z_2 \leq 10^{10}$ was found. These findings suggest that two
Reissner-Nordstr\"{o}m superextreme bodies cannot be found in
equilibrium without a strut or tension.

\subsection{One Black Hole and One Superextreme Body}

The following three different cases were found for which
Eq.~(\ref{K6eqn}) has a real root.  Each case has the configuration
illustrated in Fig.~\ref{fig2}.
\begin{description}
\item[Case A)] For $m_1=6.0,\;\; q_1=2.0, \;\; z_1=-5.0, \;\;
  m_2=-0.7,\;\; q_2=4.0,\;\; $ balance at approximately
  \cite{footnote0} $z_2=2.08$ was found. The values of $\alpha_n$ are
  $\alpha_1 = 10.3,\;\; \alpha_2 = 1.74,\;\; \alpha_3 =
  -3.11+i4.30,\;\; \alpha_4 = -3.11-i4.30$. Using
  Eqs.~(\ref{flatmass}-\ref{flatcharge}), the physical masses and
  charges are $M_1=3.95,\;\; Q_1=-0.887, \;\; M_2=1.35,\;\;
  Q_2=6.89$. Using the definitions of coordinate positions described
  in Section~\ref{non-weyl}, it was found that $Z_1=-6.03$ and $Z_2 =
  3.11$.  Thus balance has occurred for $M_1 M_2 > Q_1 Q_2$, $Q_1 Q_2
  < 0$ at a coordinate separation of ${\cal S} \equiv Z_2-Z_1 = 9.13\;
  $. Note that the parameter $m_2$ is negative but both physical
  masses are positive. The parameterizations of papers~I and~II yield
  respectively
  \begin{center}
  \begin{tabular}{rclcrcl}
   \multicolumn{3}{c}{\underline{Paper~I}} & \mbox{\ \ \ \
   \ }& \multicolumn{3}{c}{\underline{Paper~II}}  \\
   $\tilde{m}_1$ & = & $4.96$ & \mbox{}&  $\hat{m}_1$ & = & $4.36$  \\
   $\tilde{q}_1$ & = & $2.31$ & \mbox{}&  $\hat{q}_1$ & = & $-1.05$  \\
   $\tilde{m}_2$ & = & $0.34$ & \mbox{}&  $\hat{m}_2$ & = & $0.94$  \\
   $\tilde{q}_2$ & = & $3.69$ & \mbox{}&  $\hat{q}_2$ & = & $7.05$  \\
   $\tilde{z}_1$ & = & $-6.60$ & \mbox{}& $\hat{z}_1$ & = & $-6.00$ \\
   $\tilde{z}_2$ & = & $3.68$ & \mbox{}&  $\hat{z}_2$ & = & $3.08$  
   \end{tabular}
   \end{center}
   which do not agree with the integrated values of Eqs.~(\ref{mass})
   and (\ref{charge}). This demonstrates that in general none of the
   analytic parameterizations proposed, including our own, are
   suitable choices for the individual masses and charges of the
   sources.

\item[Case B)] For $m_1=9.0,\;\; q_1=3.0, \;\; z_1=-40.0, \;\;
  m_2=2.5,\;\; q_2=8.0,\;\; $ balance was found at approximately
  $z_2=34.6$. The values of $\alpha_n$ are $\alpha_1 = 48.4,\;\;
  \alpha_2 = 31.61,\;\; \alpha_3 = -34.62+i7.65,\;\; \alpha_4 =
  -34.62-i7.65$. The physical masses and charges are $M_1=8.87,\;\;
  Q_1=2.00, \;\; M_2=2.63,\;\; Q_2=9.00$. The coordinate positions
  are $ -Z_1 = 40.01,\ -Z_2 = -34.6$ Thus balance has occurred for $M_1
  M_2 > Q_1 Q_2$, $Q_1 Q_2 > 0$ at a coordinate separation of ${\cal
  S} = 74.6\; .$

\item[Case C)] For $m_1=900.0,\;\; q_1=300.0, \;\; z_1=-865.0, \;\;
  m_2=0.025,\;\; q_2=0.080,\;\; $ balance was found at approximately
  $z_2=21.581$. The values of $\alpha_n$ are $\alpha_1 = 1713.5,\;\;
  \alpha_2 = 16.474,\;\; \alpha_3 = -21.582+i0.26226,\;\; \alpha_4 =
  -21.582-i0.26226$. The physical masses and charges are
  $M_1=899.71,\;\; Q_1=298.25, \;\; M_2=0.31897,\;\; Q_2=1.8254$. The
  coordinate positions are $ -Z_1 = 865.00,\ -Z_2 = -21.582$ Thus
  balance has occurred for $Q_1 Q_2 > M_1 M_2 $, $Q_1 Q_2 > 0$ at a
  coordinate separation of ${\cal S} = 886.58\; .$
\end{description}

\subsection{Comparison with Test Particle Analysis}

Bonnor's \cite{bonnor1} examination of a test particle in the field of
a Reissner-Nordstr\"{o}m source yielded a wide variety of balance
conditions. The following cases for separation-independent equilibrium
were examined (note: $M$, $Q$ characterize the Reissner-Nordstr\"{o}m
space-time and $m$, $q$ are the test body parameters):
\begin{description}
\item[Case 1)] For $q=\epsilon m,\; Q=\eta M,\;\; \epsilon,\eta = \pm
  1$, balance occurs if $\epsilon=\eta$.
\item[Case 2)] If $m=|q|,\;\; M\neq\left| Q\right|,$ or $m\neq
  |q|,\;\; M=\left| Q\right|,$ no equilibrium is possible.
\item[Case 3)] If $mM=qQ$ but $m\neq |q|$, then no equilibrium is
  possible. 
\end{description}
Since the exact solution under study contains the Weyl-class
solution as a special case, we also find Bonnor's case~1 as a
separation-independent equilibrium condition. Case 2 or 3 cannot be
tested readily by our numerical procedure. In order to do so, one
would have to have the good fortune of correctly choosing the set
$\{m_1,m_2,q_1,q_2,z_1,z_2\}$ such that the physical masses and
charges satisfy the given conditions (i.e. $M_1=|Q_1|$ etc.). Then, to
test the dependence on separation, one would need to choose a new set
of unphysical parameters such that the proper separation changes while
the physical masses and charges remain the same.

The following separation-dependent cases were also found in
Ref.~\onlinecite{bonnor1}: 
\begin{description}
\item[Case 4)] If $\left|Q\right| > M,\;\; mM=-qQ$ and $m^2\neq q^2$
  with $qQ <0$, then an equilibrium exists at \[ r=\frac{Q^2}{2M}.\]
\item[Case 5)] If $\left|Q\right| > M,\;\; |q| < m, \;\; qQ<0$ or
\item[Case 6)] if $\left|Q\right| > M,\;\; |q| < m, \;\; qQ>0,\;\;
  qQ<mM$ then an equilibrium position exists at \[ r = \frac{Q^2\left(
    M\left(m^2-q^2\right) + q
    \sqrt{\left(m^2-q^2\right)\left(Q^2-M^2\right)}
  \right)}{m^2M^2-q^2Q^2} .\]
\item[Case 7)] If $\left|Q\right| < M,\;\; |q| > m, \;\; qQ>0,\;\;
  qQ>mM$ then an equilibrium position exists at \[ r = \frac{Q^2\left(
    M\left(m^2-q^2\right) - q
    \sqrt{\left(m^2-q^2\right)\left(Q^2-M^2\right)}
  \right)}{m^2M^2-q^2Q^2} .\]
\end{description}
Thus we have found a direct correspondence between cases A--C of the
exact solution and cases 5--7 of Bonnor's test particle analysis. The
separation dependence of cases 4--7 cannot be studied in the exact
solution using the present methods for the same reasons cases 2--3
cannot be studied. Since the separation dependence cannot be tested
using the present methods, there is little value in numerically
calculating the proper separation of the sources in cases A--C.

The physical parameters in case C could approximate a test body in a
strong gravitational field. Using these values in case 7 and
transforming from spherical coordinates to cylindrical coordinates for
a single Reissner-Nordstr\"{o}m body using the transformation (with
$\theta =0$)
\begin{eqnarray}
z&=&\left(r-M\right) \cos\theta, \nonumber \\ 
\rho &=&\sqrt{r^2-2 M r + Q^2}\, \sin\theta ,
\end{eqnarray}
Bonnor's method yields a coordinate separation of ${\cal S} = 1465.5$.
Since the separation of the bodies from these two methods are not
consistent, it would appear that case C does not sufficiently
approximate a test body.

\section{Discussion} \label{discussion}

The essential departure in the present paper from previous work is the
attempt to parameterize the solution in terms of true physical
constants of the space-time. For a static axially symmetric solution
of the Einstein-Maxwell equations, the integrals of Eq.~(\ref{mass})
and (\ref{charge}) provide the invariant parameters required for
meaningful analysis of the properties of the solution.

There are three cases of the exact solution which have not been
examined. They are an extreme body with respectively a
Reissner-Nordstr\"{o}m black hole, a superextreme body, and another
extreme body for which the solution is not of the
Weyl-class. Knowledge of the solution analytically in terms of the
physical parameters is required to analyze these cases adequately.

Ref.~\onlinecite{coopcruz} defines the terms ``undercharged'',
``overcharged'' and ``critically charged'' as follows ($i=1,2$):
\begin{eqnarray}
M_i^2 & > & Q_i^2 \ \ \ \ \ \mbox{``undercharged''} \\
M_i^2 & < & Q_i^2 \ \ \ \ \ \mbox{``overcharged''} \\
M_i^2 & = & Q_i^2 \ \ \ \ \ \mbox{``critically charged''}
\label{criticaldiscussion} 
\end{eqnarray}
For the Weyl-class, the ``lengths'' of the Weyl rods are
\cite{coopcruz} $2l_i=2\sqrt{M_i^2-Q_i^2}\, ,\; i=1,2$. If body 1 is
``critically charged'' \cite{footnote1}, then $\alpha_1=\alpha_2\;
(=d)$ since $l_1=0$ (see Fig.~\ref{fig1}). This implies that the
terminology ``critically charged'' body and ``extreme'' body may be
used interchangeably for Weyl-class solutions. If body 1 is
``undercharged'', $\alpha_1\;(= d + 2l_1)$ and $\alpha_2\;(=d)$ are
real quantities. Thus ``undercharged body'' and ``black hole'' are
synonymous terms in the Weyl-class. Finally, if body 1 is
``overcharged'', $\alpha_1\;(= d + l_1)$ and $\alpha_2\;(=d
+\bar{l_1})$ are complex conjugates. Thus the terms ``overcharged''
and ``superextreme'' are equivalent descriptions in the Weyl-class.
Unlike the Weyl-class solutions where the ``lengths'' of the Weyl rods
(real or complex) depend only upon the mass and charge of that source,
it is strongly suggested from the analysis of Section~\ref{equilibsec}
that for the general (non-Weyl-class) solution, the ``lengths'' of the
rods also depend on the mass and charge of the other source and the
distance separating the bodies as well.  It would thus be possible to
have a ``critically'' charged body (according to
Eq.~(\ref{criticaldiscussion})) for which the ``rod'' is either of
non-zero ``length'' or ``complex''.  This is important in terms of
nomenclature for describing the physics of the space-time. Since the
transition of a pair (eg.\ ($\alpha_1,\alpha_2$)) from real values to
a complex conjugate pair in Sibgatullin's \cite{sib} method defines a
differentiation of an object with a horizon to one without, it would
seem that the appropriate description would be respectively, a black
hole (horizon), ``extreme'' body (zero ``length'' Weyl rod) and
``superextreme'' body (no horizon or naked singularity) as described
in paper~I.  The descriptions ``under'', ``over'' and ``critically''
charged body should be reserved for the relations $M_i^2 > Q_i^2$,
$M_i^2 < Q_i^2$ and $M_i^2 = Q_i^2$ respectively between the
individual masses and charges. This classification scheme would
describe equilibrium conditions more precisely once all are
identified. The appropriateness of such a scheme would become apparent
when the analytic physical parameterization of the solution is known.

Bonnor's \cite{bonnor1} test particle analysis has been modified
\cite{aguirr} in such a way that the equilibrium conditions of a
charged test particle in the field of a Kerr-Newman source can be
studied. The generalization of the mathematical solution to two
spinning sources (Kerr-Newman sources) is already known
\cite{manko,chamorro}. One is able to invariantly define angular
momentum for a stationary space-time in a manner similar to
Eqs.~(\ref{mass}--\ref{charge}) because of the presence of a spacelike
Killing vector (rotational symmetry) (see Ref.~\onlinecite{dietz} and
references therein for definitions of mass and angular momentum of
stationary vacuum fields). It is unknown how the subsequent analysis
of two identical spinning bodies in paper~I based on the invariant
definitions will affect their results, if at all. However, it is clear
that the parameterization given is inadequate for the physical
analysis of the general case (non-identical bodies).

\section{Conclusions} \label{conclusion}

The solution derived in papers I, II and this paper is a
generalization of the Weyl-class double Reissner-Nordstr\"{o}m
solution. However, the analytic parameterizations presented in papers
I, II and this paper cannot in all cases be interpreted as the true
physical constants of the spacetime. The invariant physical charge for
each source is found by direct integration of Maxwell's equations. The
physical mass is invariantly defined \cite{kramer} in a manner similar
to which the charge was found. Numerical methods were used to evaluate
the invariant individual masses and charges for the axially symmetric
superposition of two Reissner-Nordstr\"{o}m bodies.  It was found that
neither the Newtonian balance condition nor critically charged bodies
are necessary for electrostatic equilibrium. The dependence of the
balance condition on the separation of the bodies is not yet known due
to the complexities involved in expressing the solution analytically
in terms of the true physical set of parameters. However, all the
balance conditions found are consistent with Bonnor's test particle
analysis. This suggests that there exist equilibrium conditions which
depend on the separation of the sources. The parameterization of this
paper is manifestly physical in the Weyl-class limit.

\acknowledgments

This research was supported, in part, by a grant from the Natural
Sciences and Engineering Research Council of Canada and a Natural
Sciences and Engineering Research Council Postgraduate Scholarship
(GPP).


\begin{table}
\caption{The values of the parameters in the parameterizations of
papers~I and~II is shown given the Weyl-Class values.}
\begin{tabular}{ccc}
Weyl-Class & Paper~I & Paper~II\\
\tableline
$m_1 = 8$ &  $\tilde{m}_1 = 14.17$ &  $\hat{m}_1 = 3.58$  \\
$q_1 = 8$ &  $\tilde{q}_1 = 14.17$ &  $\hat{q}_1 = 3.58$  \\
$m_2 = 3$ &  $\tilde{m}_2 = -3.17$ &  $\hat{m}_2 = 7.42$  \\
$q_2 = 3$ &  $\tilde{q}_2 = -3.17$ &  $\hat{q}_2 = 7.42$  \\
$z_1 = -7$ & $\tilde{z}_1 = -2.02$ & $\hat{z}_1 = -4.7$ \\
$z_2 = 7$ &  $\tilde{z}_2 = 2.02$ &  $\hat{z}_2 = 4.7$  \\
\end{tabular}
\label{table1} 
\end{table}
 

\begin{references}
\bibitem{footnote0} In cases A--C, Eq.~(\ref{K6eqn}) has been solved
to a precision of $|K|<10^{-50}$ using highly refined values of $z_2$.
\bibitem{footnote1} It should be noted that having identical roots
$\alpha_1 = \alpha_2$ is not sufficient for identifying critically
charged bodies even in the Weyl-class. The Curzon particle is such an
object with $\alpha_1 = \alpha_2$ but it is not necessarily critically
charged. See Refs.~\onlinecite{curzon} and \onlinecite{coopcruz}.
\end{references}
\end{document}